\documentclass[aps,superscriptaddress,twocolumn,color]{revtex4-2}
\usepackage{amsmath}
\usepackage{amsfonts}
\usepackage{bm}
\usepackage{graphicx}
\usepackage{array,color}
\usepackage{braket}
\usepackage[table]{xcolor}
\usepackage{xcolor}
\begin{document}
\title{Quantum many-body analysis of spin-2 bosons with two-body inelastic decay}%

\author
{
  Takeshi Takahashi and Hiroki Saito \\
  \thanks{Department of Engineering Science, University of Electro-Communications, Tokyo 182-8585, Japan}
}
\date{\today}%

\begin{abstract}
  Bose-Einstein condensates of $^{87}\mathrm{Rb}$ atoms with a
  hyperfine spin of 2 are open quantum systems,
  where the atoms are lost through two-body inelastic collisions.
  In this dissipation process, 
  a collision channel with total spin of 4 is forbidden by angular momentum conservation, which results in 
  magnetization of the atoms remaining in the condensate.
  Here, we investigate the quantum many-body properties of spin-2
  bosons that undergo two-body atomic loss.
  We show that the system finally reaches a steady state, which is a mixture of the states with maximum total spins. 
  In addition, we find that a nonclassical steady state can be
  obtained by applying and quenching the quadratic Zeeman field.
\end{abstract}
\maketitle

\section{Introduction}
%% 
%% paragraph 1
Atomic Bose-Einstein condensates (BECs) with spin degrees of freedom~\cite{Kawaguchi}
have been produced experimentally by far-off-resonant optical traps~\cite{Stamper-Kurn}.
The spin degrees of freedom enable a rich variety of ground-state phases~\cite{Stenger, Zhang, Murata} and topological excitations~\cite{Leanhardt, Kumakura}.
For example, skyrmions, monopoles, and knots have been experimentally demonstrated~\cite{Leslie, Ray1, Ray2, Hall, Lee}.
%%% Ferro, Polar, Antiferro, Broken-Axisymmetry

%% paragraph 2
Spin-2 BECs have attracted much interest
because they exhibit more diverse magnetic phases than spin-1 BECs \cite{Ciobanu,Klausen,Widera}.
However, for the atomic species used in experiments (e.g., $^{87}\mathrm{Rb}$),
the spin-2 hyperfine state has a larger energy than the spin-1 state
and is unstable against transition to the spin-1 state through binary collisions~\cite{Kuwamoto,Schmaljohann,Saito}. 
In such inelastic collisions, the hyperfine energy (e.g., $\simeq 6.8$ GHz for $^{87}\mathrm{Rb}$) transforms to 
kinetic energy and the colliding atoms escape from the trap.
Such a system with particle dissipation is an example of an open quantum system \cite{Tojo1, Tojo2, Eto}.

%% paragraph 3
In the experiment reported by Eto {\it et al.}~\cite{Eto}, magnetization was
observed to
dynamically develop in a spin-2 BEC of $^{87}\mathrm{Rb}$ atoms,
even though the spin-dependent contact interaction was not ferromagnetic.
The mechanism involved in such magnetization is spin-dependent atomic loss, 
in which two-body inelastic collisions of particles with a total spin
of 4 ($=2+2$) is prohibited
because of angular momentum conservation (i.e., only atomic pairs with different spin directions are lost).
Consequently, the remaining atoms tend to become magnetized in the same direction.

%% paragraph 4
In Ref.~\cite{Eto}, the experimental results were confirmed by numerical calculations using the mean-field approximation.
However, the mean-field picture is inappropriate for systems with a small number of particles. 
In fact, when the trap confinement is tight and all atoms occupy the same spatial wave function, 
the true ground state of a spin-1 BEC with antiferromagnetic interaction is
a symmetry-preserving state referred to as a fragmented state~\cite{Mueller, Koashi, Ueda}. 
Recently, such a quantum many-body state was experimentally observed in a BEC with $\sim 100$ atoms~\cite{Evrard1,Evrard2}.

%% paragraph 5
{\color{black}
In the present study, using the Lindblad master equation~\cite{Gardiner},
we perform a quantum many-body analysis of spin-2
bosons that undergo two-body inelastic loss.
We obtain two main results.
The first result is that the steady state reached after longtime evolution is shown to be a statistical mixture of the states with maximum total spins
(i.e., the final state is fully magnetized).
The second result is that we can obtain a nonclassical quantum many-body state (Schr\"odinger-cat-like state) as the steady state.
By applying a quadratic Zeeman field and quenching it, we can increase the probability of obtaining the nonclassical state.
We numerically solve the time evolution of the master equation for
$20$ particles and demonstrate the two aforementioned phenomena.
}

%% paragraph 6
The remainder of this paper is organized as follows.
First, we formulate the present open quantum system in Sec.~\ref{formulation}
and describe the dynamics of the system using the Lindblad master equation.
In Sec.~\ref{theory}, we focus on the steady states of this equation
and analyze their general properties.
In Sec.~\ref{results}, we present the results of numerical simulations
of the dynamics of the system.
The conclusions to this study are provided in Sec.~\ref{conclusions}.

\section{Formulation of the problem}\label{formulation}

\subsection{Master equation with atomic loss}
%% 
%% paragraph 1
We consider spin-2 bosonic atoms confined in a trap potential $V(\bm{r})$. 
The field operators $\hat{\psi}_m(\bm{r})$ annihilate atoms in the magnetic sublevels
$m = -2, -1, 0, 1, 2$, which satisfy the bosonic commutation relations $[\hat{\psi}_m(\bm{r}), \hat{\psi}_{m^\prime}^\dagger(\bm{r}^\prime)] = \delta(\bm{r} - \bm{r}^\prime)\delta_{m, m^\prime}$. 
The non-interacting component of the Hamiltonian is given by
\begin{equation}
  \begin{aligned}
  \label{H0}
    \hat{H}_0 = \int \sum_{m = -2}^2 \left\{ \hat{\psi}^\dagger_m(\bm{r}) \left[ -\frac{\hbar^2 \nabla^2}{2M} + V(\bm{r}) \right] \hat{\psi}_m(\bm{r}) \right. \\
     + qm^2 \hat{\psi}^\dagger_m(\bm{r}) \hat{\psi}_m(\bm{r}) \bigg{\}} d\bm{r},
  \end{aligned}
\end{equation}
where $M$ is the mass of an atom and $q$ is the quadratic Zeeman coefficient.
Here, we ignore the linear Zeeman effect because it can be eliminated by the unitary transformation.
{\color{black}
The quadratic Zeeman term is realized by applying an external magnetic field.
}
%%
%% paragraph 2
The interaction term $\hat{H}_{\mathrm{int}}$ that represents the elastic collisions is written as
\begin{equation}
\label{Hint}
  \hat{H}_{\mathrm{int}} = \int \sum_{\mathcal{F} = 0, 2, 4} \frac{g_\mathcal{F}}{2} \sum_{\mathcal{M} = -\mathcal{F}}^\mathcal{F} \hat{A}^\dagger_{\mathcal{FM}}(\bm{r}) \hat{A}_{\mathcal{FM}}(\bm{r}) d\bm{r},
\end{equation}
where $\mathcal{F}$ is the total spin of the colliding atoms
and $g_\mathcal{F} = 4\pi\hbar^2 a_\mathcal{F} / M$ with $a_\mathcal{F}$
being the $s$-wave scattering length of the collision channel with $\mathcal{F}$.
The operator $\hat{A}_{\mathcal{FM}}$ is defined as
\begin{equation}
  \hat{A}_{\mathcal{FM}}(\bm{r}) = \sum_{m, m^\prime = -2}^2 C_{m, m^\prime}^{\mathcal{FM}} \hat{\psi}_m(\bm{r})\hat{\psi}_{m^\prime}(\bm{r}),
\end{equation}
where $C_{m,m'}^{\mathcal{FM}}$ is the Clebsch-Gordan coefficient.
The Hamiltonian for the total system is given by $\hat{H} = \hat{H}_0 + \hat{H}_{\mathrm{int}}$.

%% paragraph 3
Next, we consider two-body inelastic collisions, in which a transition
from a hyperfine spin-2 state to a hyperfine spin-1 state occurs.
The energy difference between hyperfine spin-2 and spin-1 states for $^{87}\mathrm{Rb}$ is $\simeq 6.8$ GHz, 
and this energy is transferred to the kinetic energy of the colliding
atoms, causing them to escape from the trap.
The dynamics of such an open quantum system with atomic loss can be described by the density operator $\hat{\rho}(t)$ as
\begin{equation}
\label{open_quantum_equation}
  \frac{d \hat{\rho}(t)}{d t} = \frac{1}{i\hbar}[\hat{H}, \hat{\rho}(t)] + \left(\frac{d \hat{\rho}(t)}{d t}\right)_{\mathrm{loss}},
\end{equation}
where $(d \hat{\rho}(t) / d t)_{\mathrm{loss}}$
represents the atomic loss due to inelastic collisions. 
Assuming that the atoms escaping from the trap do not interact with other atoms,
the atomic loss is described by a Markov process. 
In this case, $(d \hat{\rho}(t) / d t)_{\mathrm{loss}}$ 
in Eq.~(\ref{open_quantum_equation}) is given in the Lindblad form as \cite{Tojo2}
\begin{equation}
\begin{aligned}
\label{dissipation_term}
  \left(\frac{d \hat{\rho}(t)}{d t}\right)_{\mathrm{loss}} = \int \sum_{\mathcal{F} = 0, 2}\frac{b_\mathcal{F}}{4}\sum_{\mathcal{M} = -\mathcal{F}}^\mathcal{F} \left[ 2\hat{A}_{\mathcal{FM}}(\bm{r})\hat{\rho}(t)\hat{A}_{\mathcal{FM}}^\dagger(\bm{r}) \right. \\
  \left. - \hat{A}_{\mathcal{FM}}^\dagger(\bm{r}) \hat{A}_{\mathcal{FM}}(\bm{r})\hat{\rho}(t) - \hat{\rho}(t)\hat{A}_{\mathcal{FM}}^\dagger(\bm{r}) \hat{A}_{\mathcal{FM}}(\bm{r}) \right] d\bm{r}, 
\end{aligned}
\end{equation}
where $b_\mathcal{F}$ is the loss coefficient for a collision channel with total spin $\mathcal{F}$.
Notably, for $\mathcal{F} = 4$, inelastic collisions are forbidden
(i.e., $b_4 = 0$) because of the conservation of spin angular momentum. 
The first term in the square brackets in Eq.~(\ref{dissipation_term}) is called
a jump term, which eliminates two particles from the system.

\subsection{Previous results}

%%
%% paragraph 1
Here, we briefly review the results of a previous study on a spin-2 $^{87}\mathrm{Rb}$ BEC with particle dissipation~\cite{Eto}.
In this previous study, all atoms were initially in the $m = 0$ state and the subsequent dynamics was investigated.
It was observed that the system eventually became fully magnetized in the transverse direction
despite the ferromagnetic state being energetically unfavorable~\cite{Widera}.
This is due to the spin-dependent inelastic collisions ($b_4 = 0$);
that is, atoms with different spin directions are selectively lost.

%%
%% paragraph 2
A numerical analysis was also performed in Ref.~\cite{Eto}, in which Eq.~(\ref{open_quantum_equation}) was
reduced to the dissipative Gross-Pitaevskii equation.
In the mean-field approximation, 
the macroscopic wave function $\psi_m(\bm{r}, t) \equiv \mathrm{Tr}[\hat{\psi}_m(\bm{r})\hat{\rho}(t)]$ obeys~\cite{Tojo2}
\begin{equation}
\label{GP}
\begin{aligned}
  & i\hbar\frac{\partial \psi_m(\bm{r}, t)}{\partial t} = \left[ -\frac{\hbar^2 \nabla^2}{2M} + V(\bm{r}) + qm^2 \right]\psi_m(\bm{r}, t) \\
  & + \sum_{\substack{\mathcal{F} = 0, 2, 4 \\ \mathcal{M} = -\mathcal{F}, \dots, \mathcal{F}}} \tilde{g}_\mathcal{F} \sum_{\substack{i, j, k = -2}}^2 C_{m, i}^{\mathcal{FM}}C_{j, k}^{\mathcal{FM}}\psi_{i}^\star(\bm{r}, t)\psi_{j}(\bm{r}, t)\psi_{k}(\bm{r}, t),
\end{aligned}
\end{equation}
where $\tilde{g}_\mathcal{F} = g_\mathcal{F} - i\hbar b_\mathcal{F} / 2$ for $\mathcal{F} = 0, 2$, and $\tilde{g}_4 = g_4$.
In Ref.~\cite{Eto}, the numerical solutions of Eq.~(\ref{GP}) were
found to agree well with the experimental results.
However, the mean-field approximation is only valid for a large number of atoms and we do not use Eq.~(\ref{GP}) in the present study.

\subsection{Single-mode approximation}

%%
%% paragraph 1
We consider a situation in which atoms are tightly trapped and the size of the atomic cloud is much smaller than the spin healing length.
In this case, all atoms occupy the same spatial wave function with a uniform spin state and we can use the single-mode approximation. 
The field operators can be rewritten as
\begin{equation}
\label{SMA}
  \hat{\psi}_m(\bm{r}) \simeq \hat{a}_m \psi_{\mathrm{SMA}}(\bm{r}),
\end{equation}
where $\hat{a}_m$ are bosonic operators and the wave function $\psi_{\mathrm{SMA}}(\bm{r})$ 
is the ground state of $[-\hbar^2 \nabla^2 /(2M) + V] \psi_{\rm SMA} = \varepsilon_0 \psi_{\rm SMA}$ satisfying
the normalization condition $\int |\psi_{\mathrm{SMA}}(\bm{r})|^2 d\bm{r} = 1$.
Substituting Eq.~(\ref{SMA}) into Eqs.~(\ref{H0}) and (\ref{Hint}), we obtain
\begin{equation}
\label{HofSMA}
\begin{aligned}
  \hat{H} = &\varepsilon_0 \hat{N} + q \sum_{m = -2}^2 m^2 \hat{a}^\dagger_m \hat{a}_m \\ 
  &+ \sum_{\mathcal{F} = 0, 2, 4} \frac{g_\mathcal{F}}{2V^\mathrm{eff}} \sum_{\mathcal{M} = -\mathcal{F}}^\mathcal{F} \hat{A}^\dagger_{\mathcal{FM}} \hat{A}_{\mathcal{FM}},
\end{aligned}
\end{equation}
where $\hat{N} = \sum_m \hat{a}_m^\dagger \hat{a}_m$, $V^{\rm eff} = \left[\int |\psi_{\mathrm{SMA}}(\bm{r})|^4 d\bm{r}\right]^{-1}$, and
\begin{equation}
  \hat{A}_{\mathcal{FM}} = \sum_{m, m^\prime = -2}^2 C_{m, m^\prime}^{\mathcal{FM}} \hat{a}_m\hat{a}_{m^\prime}.
\end{equation}
{\color{black}
Similarly, substituting Eq.~(\ref{SMA}) into Eq.~(\ref{dissipation_term}) gives the single-mode approximation of the particle loss terms.
We thus obtain the master equation for the single-mode approximation as
\begin{equation}
\label{open_quantum_equation_ofSMA}
  \begin{aligned}
  \frac{d\hat{\rho}(t)}{dt} &= \frac{1}{i\hbar} [\hat{H}, \hat{\rho}(t)] \\
  & + \sum_{\mathcal{F} = 0, 2}\frac{b_\mathcal{F}}{4 V^\mathrm{eff}} 
  \sum_{\mathcal{M} = -\mathcal{F}}^\mathcal{F} \left[ 2\hat{A}_{\mathcal{FM}}\hat{\rho}(t)\hat{A}_{\mathcal{FM}}^\dagger \right. \\
  &\left. - \hat{A}_{\mathcal{FM}}^\dagger \hat{A}_{\mathcal{FM}}\hat{\rho}(t) - \hat{\rho}(t)\hat{A}_{\mathcal{FM}}^\dagger \hat{A}_{\mathcal{FM}} \right], 
  \end{aligned}
\end{equation}
where $\hat{H}$ is given by Eq.~(\ref{HofSMA}).
}

%% paragraph 2
The energy eigenstates of $\hat{H}$ in Eq.~(\ref{HofSMA}) are expressed in the form $|N_0, N_S, F, F_z\rangle$ for $q=0$,
where $N_S$ is the number of particles forming singlet pairs, {\color{black}$N_0$ is the number of remaining particles, }
$F$ is the total spin, and $F_z$ is the magnetic quantum number \cite{Koashi}. 
{\color{black}
The total number of particles is $N = N_0 + 2N_S$.
}
The energy eigenvalue is given by
\begin{equation}
  \label{eigenvalue}
\begin{aligned}
  E = &\varepsilon_0 N + \frac{c_0}{2V^\mathrm{eff}} N (N-1) \\ + &\frac{c_1}{2V^\mathrm{eff}}[F(F+1) - 6N] + \frac{c_2}{5V^\mathrm{eff}}N_S(N + N_0 + 3), 
\end{aligned}
\end{equation}
where $c_0, c_1$, and $c_2$ are defined as 
\begin{equation}
  \begin{aligned}
  c_0 &= \frac{4g_2 + 3g_4}{7},\\
  c_1 &= \frac{g_4 - g_2}{7}, \\
  c_2 &= \frac{7g_0 -10 g_2 + 3g_4}{7}.
  \end{aligned}
\end{equation}
We note that the energy eigenstates $|N_0,N_S,F,F_z\rangle$ are independent of the values of the interaction parameters $g_0$, $g_2$, and $g_4$.

{\color{black}
\section{STEADY STATES REACHED AFTER LONG TIME}\label{theory}

%% paragraph 1
This section provides a general discussion of the steady state for the master equation (\ref{open_quantum_equation_ofSMA}) with a single-mode approximation. 
The steady states satisfy 
\begin{equation}
\label{steady_state_condition}  
\frac{d \hat{\rho}(t)}{d t} = 0.
\end{equation}
In Sec.~\ref{two_particle_solution}, we investigate the form of the steady state for a system with an initial particle number of two;
in Sec.~\ref{general_solution_steady_states}, we extend the discussion to general particle numbers.

\subsection{Two-particle solution}\label{two_particle_solution}

%% paragraph 2
As a simple toy model, we consider an initial state in which two atoms occupy the $m = 0$ state:
\begin{equation}
  \label{initial_state_2particles}
  |0, 0, 2, 0, 0 \rangle = \frac{1}{\sqrt{2}}(\hat{a}_0^\dagger)^2 |\mathrm{vac}\rangle,
\end{equation}
where $|\mathrm{vac}\rangle$ is the vacuum state.
We solve the master equation (\ref{open_quantum_equation_ofSMA}) and obtain the steady state for $t \to \infty$. 

%% paragraph 3
We abbreviate the relevant energy eigenstates as
\begin{equation}
  \label{basis_2particles}
\begin{aligned}
|0\rangle &\equiv |\mathrm{vac}\rangle, \\
|1\rangle &\equiv |N_0=0, N_S=1, F=0, F_z=0\rangle, \\
|2\rangle &\equiv |N_0=2, N_S=0, F=2, F_z=0\rangle, \\
|3\rangle &\equiv |N_0=2, N_S=0, F=4, F_z=0\rangle.
\end{aligned}
\end{equation}
The initial state in Eq.~(\ref{initial_state_2particles}) can be expanded by these energy eigenstates as
%% we can rewrite Eq.~(\ref{initial_state_2particles}) as a superposition of Eq.~(\ref{basis_2particles}) as
\begin{equation}
  \label{initial_state_expanded}
|0, 0, 2, 0, 0\rangle = \frac{1}{\sqrt{35}}\left(\sqrt{7}|1\rangle  - \sqrt{10} |2\rangle+ 3\sqrt{2} |3\rangle\right).
\end{equation}
We write the time evolution of the density operator as 
\begin{equation}
  \hat{\rho}(t) = \sum_{i, j = 0}^3 c_{ij}(t) |i\rangle \langle j|.
\end{equation}
Applying $\hat{A}_\mathcal{FM}$ to the states $|i \rangle$, we have
$\hat{A}_{00}|1\rangle = \hat{A}_{20}|2\rangle = \sqrt{2} |0\rangle$ with all other combinations being zero;
their Hermitian conjugates give $\hat{A}_{00}^\dagger|0\rangle = \sqrt{2} |1\rangle$ and $ \hat{A}_{20}^\dagger|0\rangle = \sqrt{2} |2\rangle$.
Consequently, 
Eq.~(\ref{open_quantum_equation_ofSMA}) is expressed in the form
\begin{equation}
  \label{time_evolution_2particles}
  \begin{aligned}
\frac{d \hat{\rho}(t)}{d t}=&\frac{1}{i\hbar}\sum_{i, j=0}^3 \left[E_i - E_j - i(D_i + D_j)\right]c_{ij}(t)|i\rangle \langle j|\\
&+2D_1 c_{11}(t) |0\rangle \langle 0| + 2D_2 c_{22}(t) |0\rangle \langle 0|,
  \end{aligned}
\end{equation}
where $E_i$ and $D_i$ are $E_0 = D_0 = 0, E_1 = g_0/V^\mathrm{eff}, D_1 = b_0 / (2V^\mathrm{eff}), 
E_2 = g_2/V^\mathrm{eff}, D_2 = b_2 / (2V^\mathrm{eff})$, $E_3 = g_4/V^\mathrm{eff}$, and $D_3 = 0$.

%% paragraph 4
We solve Eq.~(\ref{time_evolution_2particles}) for the initial condition of $\hat{\rho}(0) = |0, 0, 2, 0, 0 \rangle \langle 0, 0, 2, 0, 0 |$.
For $i \neq j$, we have 
\begin{equation}
  \frac{d c_{ij}(t)}{d t} = \frac{1}{i\hbar}[E_i - E_j - i(D_i + D_j)]c_{ij}(t),
\end{equation}
giving $c_{ij}(t) = c_{ij}(0) \exp\{[E_i - E_j - i(D_i + D_j)]t / (i\hbar)\}$.
Since $D_i + D_j > 0$ for $i \neq j$, the off-diagonal coefficients become $c_{ij}(t) \to 0$ as $t \to \infty$.
For the diagonal elements $c_{ii}$, Eq.~(\ref{time_evolution_2particles}) gives
\begin{equation}
  \begin{aligned}
  \frac{d c_{00}(t)}{d t} &= 2D_1 c_{11}(t) + 2D_2 c_{22}(t), \\
  \frac{d c_{11}(t)}{d t} &= -2D_1 c_{11}(t), \\
  \frac{d c_{22}(t)}{d t} &= -2D_2 c_{22}(t), \\
  \frac{d c_{33}(t)}{d t} &= 0,
  \end{aligned}
\end{equation}
which are solved to give
%% and their solutions are given by 
\begin{equation}
  \begin{aligned}
  c_{00}(t) &= c_{11}(0)(1 - e^{-2D_1 t}) + c_{22}(0)(1 - e^{-2D_2 t}), \\
  c_{11}(t) &= c_{11}(0) e^{-2D_1 t}, \\
  c_{22}(t) &= c_{22}(0) e^{-2D_2 t}, \\
  c_{33}(t) &= c_{33}(0).
  \end{aligned}
\end{equation}
In the limit of $t \to \infty$, $c_{11}(t) \rightarrow 0$ and $c_{22}(t) \rightarrow 0$, whereas $c_{33}(t)$ maintains its initial value.
Thus, we obtain the steady state as
\begin{equation}
  \label{steady_state_2particles}
\hat{\rho}(t) \rightarrow \frac{17}{35} |0\rangle \langle0| + \frac{18}{35} |3\rangle \langle 3|~~~~(t \rightarrow \infty).
\end{equation}
Other than the vacuum state $|0\rangle$, this steady state only contains the state
$|3\rangle = |N_0=2, N_S=0, F=4, F_z=0\rangle$ that has the maximum total spin $F = 2N = 4$.
This is because particle loss occurs only through the ${\cal F} = 0$ and ${\cal F} = 2$ channels.
The maximally polarized state therefore does not undergo particle loss and survives in the steady state.
The above simple analysis for two particles implies that for general
numbers of particles also, the steady state only contains the states
with maximum total spins.
In the next subsection, we will show that this is the case.
}
{\color{black}
\subsection{General forms of steady states}\label{general_solution_steady_states}

%% paragraph 1
In this subsection, we prove
that the steady state satisfying Eq.~(\ref{steady_state_condition}) is a mixture of energy eigenstates with the maximum total spin.
In the following, we denote the density operator for the $N$-particle subspace as $\hat{\rho}_N$ and express the density operator for the total Hilbert space as
\begin{equation}
\label{rho_total_17.5}
  \hat{\rho}(t) = \oplus_{N=0}^{N_\mathrm{ini}}\hat{\rho}_N(t),
\end{equation}
where $N_\mathrm{ini}$ is the maximum number of atoms contained in the initial state.

%% paragraph 2
First, we consider the case of $q = 0$. We abbreviate the states with the maximum total spin $F = 2N$ as
\begin{equation}
  \label{maxF_general}
  |F = 2N, F_z \rangle \equiv |N_0 = N, N_S = 0, F = 2N, F_z\rangle,
\end{equation}
where $-2N \leq F_z \leq 2N$.
We define a density operator that only consists of the states with maximum total spin as
\begin{equation}
\label{rho_N_general}
  \hat\rho^\mathrm{ss}_N \equiv \sum_{F_z, F_z'} c_{F_z F_z'} |F=2N, F_z\rangle \langle F=2N, F_z'|,
\end{equation}
where $c_{F_z F_z'}$ is an element of a positive semidefinite Hermitian matrix.
Using Eq.~(\ref{rho_N_general}), we define the density operator as
\begin{equation}
  \label{rho_ss_general}
\hat{\rho}^\mathrm{ss} \equiv \oplus_{N=0}^{N_\mathrm{ini}} \hat{\rho}^\mathrm{ss}_N.
\end{equation}
We will prove that Eq.~(\ref{rho_ss_general}) is a necessary and sufficient condition for a steady-state solution of Eq.~(\ref{open_quantum_equation_ofSMA}) satisfying
\begin{equation}
  \label{steady_state_equation}
  \begin{aligned}
\frac{1}{i\hbar}&\left[\hat{H}, \hat{\rho}^\mathrm{ss}\right] 
+ \sum_{\mathcal{F} = 0, 2}\frac{b_\mathcal{F}}{4 V^\mathrm{eff}} \sum_{\mathcal{M} = -\mathcal{F}}^\mathcal{F} \left( 2\hat{A}_{\mathcal{FM}}\hat{\rho}^\mathrm{ss}\hat{A}_{\mathcal{FM}}^\dagger \right. \\
& \left. - \hat{A}_{\mathcal{FM}}^\dagger \hat{A}_{\mathcal{FM}}\hat{\rho}^\mathrm{ss} - \hat{\rho}^\mathrm{ss}\hat{A}_{\mathcal{FM}}^\dagger \hat{A}_{\mathcal{FM}} \right) = 0.
  \end{aligned}
\end{equation}
}
%% paragraph 3
First, we show that Eq.~(\ref{rho_ss_general}) is a sufficient condition for Eq.~(\ref{steady_state_equation}).
The commutator in Eq.~(\ref{steady_state_equation}) vanishes,
\begin{equation}
  \label{commutation_zero_general}
\frac{1}{i\hbar} \left[\hat{H}, \hat{\rho}^\mathrm{ss}\right] = 0, 
\end{equation}
since the bases contained in Eq.~(\ref{rho_N_general}) are eigenstates of $\hat{H}$ with the same eigenvalues.
The state in Eq.~(\ref{maxF_general}) is expressed as \cite{Koashi}
\begin{equation}
  \label{maxF_explicit}
  \begin{aligned}
  |F = 2N, F_z\rangle = Z^{-1/2}(\hat{F}_-)^{\Delta F} (\hat{a}^\dagger_2)^{N}|\mathrm{vac}\rangle,
  \end{aligned}
\end{equation}
where $\hat{F}_-=2(\hat{a}^\dagger_1 \hat{a}_2 + \hat{a}_{-2}^\dagger \hat{a}_{-1}) + \sqrt{6}(\hat{a}^\dagger_0\hat{a}_1 + \hat{a}^\dagger_{-1}\hat{a}_0)$, 
$\Delta F = 2N - F_z$, and $Z = N!(4N)!\Delta F! / (4N-\Delta F)!$.
From the form of $\hat{A}_{00} = (\hat{a}_0 \hat{a}_0 - 2\hat{a}_1\hat{a}_{-1} + 2\hat{a}_2\hat{a}_{-2})/\sqrt{5}$, we find $\hat{A}_{00} (\hat{a}^\dagger_2)^{N}|\mathrm{vac}\rangle = 0$ and $[\hat{A}_{00}, \hat{F}_-]=0$,
which gives $\hat{A}_{00} (\hat{F}_-)^{\Delta F}(\hat{a}^\dagger_2)^{N}|\mathrm{vac}\rangle = 0$.
Similarly, $\hat{A}_{2\mathcal{M}}$ satisfies $\hat{A}_{2\mathcal{M}} (\hat{a}^\dagger_2)^{N}|\mathrm{vac}\rangle = 0$. 
From the commutation relations between $\hat{A}_{2\mathcal{M}}$ and $\hat{F}_-$,
\begin{equation}
\label{commutation_AF}
\begin{aligned}
&\left[ \hat{A}_{2,2}, \hat{F}_- \right] = 0, \\
&\left[ \hat{A}_{2,1}, \hat{F}_- \right] = 2 \hat{A}_{2, 2}, \\
&\left[ \hat{A}_{2,0}, \hat{F}_- \right] = \sqrt{6} \hat{A}_{2, 1}, \\
&\left[ \hat{A}_{2,-1}, \hat{F}_- \right] = 2\sqrt{6}\hat{A}_{2, 0}, \\
&\left[ \hat{A}_{2,-2}, \hat{F}_- \right] = 4 \hat{A}_{2, -1}, \\
\end{aligned}
\end{equation}
we obtain $\hat{A}_{2\mathcal{M}}(\hat{F}_-)^{\Delta F}(\hat{a}^\dagger_2)^{N}|\mathrm{vac}\rangle = 0$.
Thus, we have shown that
\begin{equation}
  \label{A_FM_vanish}
\hat{A}_\mathcal{FM} |F=2N, F_z\rangle = 0
\end{equation}
for $\mathcal{F} = 0$ and $\mathcal{F} = 2$, clearly indicating that the second
term on the left-hand side of Eq.~(\ref{steady_state_equation}) also vanishes,
\begin{equation}
  \label{dissipation_zero_general}
  \begin{aligned}
&  \sum_{\mathcal{F} = 0, 2}\frac{b_\mathcal{F}}{4 V^\mathrm{eff}} \sum_{\mathcal{M} = -\mathcal{F}}^\mathcal{F} \left( 2\hat{A}_{\mathcal{FM}}\hat{\rho}^\mathrm{ss}\hat{A}_{\mathcal{FM}}^\dagger \right. \\
& \left. - \hat{A}_{\mathcal{FM}}^\dagger \hat{A}_{\mathcal{FM}}\hat{\rho}^\mathrm{ss} - \hat{\rho}^\mathrm{ss}\hat{A}_{\mathcal{FM}}^\dagger \hat{A}_{\mathcal{FM}} \right) = 0.
  \end{aligned}
\end{equation}

%% paragraph 4
{\color{black}
Next, we prove that Eq.~(\ref{rho_ss_general}) is a necessary condition for Eq.~(\ref{steady_state_equation}).
In the steady state satisfying Eq.~(\ref{steady_state_condition}),
the time derivative of the average particle number $d\langle \hat{N}\rangle / dt$ must be zero.
Using the master equation (\ref{open_quantum_equation_ofSMA}) and the commutation relation
\begin{equation}
[\hat{N}, \hat{A}_\mathcal{FM}] = - 2\hat{A}_\mathcal{FM},
\end{equation}
we can express the time derivative of the average particle number as
\begin{equation}
\label{derivative_N}
\begin{aligned}
\frac{d\langle \hat{N}\rangle}{dt} &= \mathrm{Tr}\left[\hat{N}\frac{d \hat{\rho}(t)}{dt}\right]\\
&=-\mathrm{Tr}\left[ \sum_{\mathcal{F} = 0, 2}\frac{b_\mathcal{F}}{V^\mathrm{eff}}\sum_{\mathcal{M} = -\mathcal{F}}^\mathcal{F} \hat{A}_\mathcal{FM}\hat{\rho}\hat{A}_\mathcal{FM}^\dagger \right] \\
&=-\sum_{\mathcal{F} = 0, 2}\frac{b_\mathcal{F}}{V^\mathrm{eff}}\sum_{\mathcal{M} = -\mathcal{F}}^\mathcal{F} \mathrm{Tr} \left[\hat{A}_\mathcal{FM}^\dagger \hat{A}_\mathcal{FM}\hat{\rho} \right] \\
&= 0.
\end{aligned}
\end{equation}
%% %% paragraph 5
From the positive definiteness of $\hat{A}_\mathcal{FM}^\dagger \hat{A}_\mathcal{FM}$, each term of the sum in the third line of Eq.~(\ref{derivative_N}) must vanish.
Equation (\ref{derivative_N}) therefore leads to
\begin{equation}
  \label{matrix_condition}
\hat{A}_\mathcal{FM} \hat{\rho} \hat{A}_\mathcal{FM}^\dagger = 0, 
\end{equation}
for $\mathcal{F} = 0, 2$.
}

%% paragraph 6
In general, the density operator $\hat{\rho}$ is written as a mixture of pure states $\{|\psi_k\rangle\}_{k = 1, 2, \dots, K}$ as
\begin{equation}
  \label{diagonalize}
\hat{\rho} = \sum_{k = 1}^K P_k|\psi_k\rangle\langle\psi_k|,
\end{equation}
where $P_k$ is positive.
When the density operator satisfying Eq.~(\ref{matrix_condition}) is expressed by Eq.~(\ref{diagonalize}), 
\begin{equation}
  \label{vector_condition}
  \begin{aligned}
\hat{A}_\mathcal{FM} |\psi_k\rangle &= 0~~~~(\mathcal{F} = 0, 2)
  \end{aligned}
\end{equation}
must hold for each pure state $|\psi_k\rangle$.
The states in Eq.~(\ref{maxF_general}) satisfy Eq.~(\ref{vector_condition}), as shown in Eq.~(\ref{A_FM_vanish}).
We can show that no other states satisfy Eq.~(\ref{vector_condition}) as follows.
Equation~(\ref{vector_condition}) leads to
\begin{equation}
\label{vanish_condition}
\sum_{\mathcal{F}=0, 2}\frac{b_\mathcal{F}}{2V^\mathrm{eff}} \sum_{\mathcal{M} = 
-\mathcal{F}}^\mathcal{F} \hat{A}^\dagger_\mathcal{FM} \hat{A}_\mathcal{FM} |\psi_k\rangle = 0.
\end{equation}
Since the operator on the left-hand side has the same form as the Hamiltonian in Eq.~(\ref{HofSMA}) with $\varepsilon_0 = g_4 = q = 0$, the eigenvalue $\lambda$ is given by
\begin{equation}
\label{maxF_eigenvalue}
\begin{aligned}
\lambda = &\frac{b_2}{14V^\mathrm{eff}}[2N(2N + 1) - F(F+1)] \\ + &\frac{7b_0 - 10 b_2}{35V^\mathrm{eff}}N_S(N + N_0 + 3),
\end{aligned}
\end{equation}
from Eq.~(\ref{eigenvalue}).
The quantum numbers $N_S$ and $F$ that give $\lambda = 0$ are only
$N_S = 0$ and $F = 2N$; that is, Eq.~(\ref{vanish_condition}) is satisfied only by superpositions of the states in Eq.~(\ref{maxF_general}).
Therefore, Eq.~(\ref{vector_condition}) is satisfied only by the states in Eq.~(\ref{maxF_general}).
Thus, Eq.~(\ref{rho_ss_general}) is a necessary condition for Eq.~(\ref{steady_state_equation}).
In conclusion, Eq.~(\ref{steady_state_condition}) for $q = 0$ is satisfied if and only if the density operator has the form of Eqs.~(\ref{rho_N_general}) and (\ref{rho_ss_general}).

%% paragraph 7
Finally, we consider the steady state for $q \neq 0$.
{\color{black}
The above proof is valid also for $q \neq 0$ if all of the bases used in Eq.~(\ref{rho_N_general}) are eigenstates of $\hat{H}$ with the same eigenvalue.
}However, because of the quadratic Zeeman term in the Hamiltonian,
some of the states in Eq.~(\ref{maxF_general}) are not eigenstates of the Hamiltonian;
for these states, Eq.~(\ref{commutation_zero_general}) does not hold.
Among the states in Eq.~(\ref{maxF_general}), only the states
\begin{equation}
\label{states_F2N_31}
\begin{aligned}
|F = 2N, F_z &= -2N\rangle  &&= \frac{(\hat{a}^\dagger_{-2})^N}{\sqrt{N!}} |\mathrm{vac}\rangle, \\ 
|F = 2N, F_z &= 2N\rangle &&=\frac{(\hat{a}^\dagger_{2})^N }{\sqrt{N!}}|\mathrm{vac}\rangle,
\end{aligned}
\end{equation}
and
\begin{equation}
\label{states_F2N_32}
\begin{aligned}
|F = 2N, F_z &= -2N + 1\rangle &&= \frac{\hat{a}^\dagger_{-1}(\hat{a}^\dagger_{-2})^{N-1}}{\sqrt{(N-1)!}} |\mathrm{vac}\rangle, \\
|F = 2N, F_z &= 2N - 1\rangle &&= \frac{\hat{a}^\dagger_1(\hat{a}^\dagger_{2})^{N-1}}{\sqrt{(N-1)!}} |\mathrm{vac}\rangle, \\
\end{aligned}
\end{equation}
are eigenstates of the Hamiltonian with eigenvalues $E_{q=0} + 4qN$ for Eq.~(\ref{states_F2N_31}) 
and $E_{q=0} + 4q(N-1) + q$ for Eq.~(\ref{states_F2N_32}), where $E_{q=0}$ is the eigenvalue of the Hamiltonian for $q = 0$.
Thus, if the density operator in Eq.~(\ref{rho_N_general}) is constructed from the states 
in Eqs.~(\ref{states_F2N_31}) and (\ref{states_F2N_32}) such that Eq.~(\ref{commutation_zero_general}) is satisfied, we obtain a steady state for $q \neq 0$.

\section{NUMERICAL RESULTS}\label{results}

%% table
\begin{table}[!b]
\begin{center}
\caption{Interaction and dissipation parameters $g_\mathcal{F}$ and $b_\mathcal{F}$ of $^{87}\mathrm{Rb}$, taken from Refs.~\cite{Ciobanu} and \cite{Tojo2}, respectively.}\label{table1}
\begin{tabular}{cc|cc}
  Interaction & $10^{-11} \mathrm{cm}^3 / \mathrm{s}$ & Loss & $10^{-11} \mathrm{cm}^3 / \mathrm{s}$\\ \hline
  $g_0 / \hbar$ & 4.34 & $b_0$ & 0.0099\\ \hline
  $g_2 / \hbar$ & 4.59 & $b_2$ & 0.0243 \\ \hline
  $g_4 / \hbar$ & 5.15 & $b_4$ & - \\ \hline
\end{tabular}
\end{center}
\end{table}

%% paragraph 1
We study the time evolution of the system by numerically solving the master equation~(\ref{open_quantum_equation})
using the fourth-order Runge-Kutta method.
The initial state is a Fock state 
\begin{equation}
\label{initial_state}
|\Psi_\mathrm{ini}\rangle = \frac{(\hat{a}^\dagger_0)^{N_\mathrm{ini}}}{\sqrt{N_\mathrm{ini}!}} |\mathrm{vac}\rangle,
\end{equation}
occupying the $m = 0$ state, as in the experiment in Ref.~\cite{Eto}.
Here, we restrict ourselves to $N_\mathrm{ini} = 20$, which requires $\sim 10^9$ matrix elements to represent the density operator.
We consider an atomic gas of $^{87}\mathrm{Rb}$ confined in an isotropic harmonic potential with a trap frequency $\omega / (2 \pi) = 2~\mathrm{kHz}$. 
We use the single-particle ground-state wave function for the single-mode approximation, and the effective volume
is given by $V^{\rm eff} = \left[\int |\psi_{\mathrm{SMA}}(\bm{r})|^4 d\bm{r}\right]^{-1} \simeq 2.2 \times 10^{-13}~\mathrm{cm}^3$. 
The interaction and dissipation parameters are set to the experimentally measured values given in Table \ref{table1}. 
For these interaction parameters, the ground state of the Hamiltonian for $q = 0$ is the state where all atoms form singlet pairs \cite{Koashi}.

%% paragraph 2
As the system evolves over time,
the total density operator for the system has the form $\hat{\rho}=\oplus_{N = 0}^{20} \hat{\rho}_N$, where $N$ are even integers.
The expectation value of an operator $\hat{A}$ is given by $\langle \hat{A} \rangle = \mathrm{Tr}[\hat{A} \hat{\rho}] / \mathrm{Tr}[\hat{\rho}]$.
%% and that of $\hat{A}$ within the $N$-particle subspace is defined by $\langle \hat{A}\rangle_N = \mathrm{Tr}[\hat{A}\hat{\rho}_N] / \mathrm{Tr}[\hat{\rho}_N]$.
The spin operator in the $j$-direction ($j = x, y, z$) is expressed as 
$\hat{F}_j = \sum_{m, m^\prime = -2}^2 [f_j]_{m, m^\prime} \hat{a}_m^\dagger \hat{a}_{m^\prime}$,
where $f_j$ are spin-2 matrices.
For the present initial state, $\langle \hat{F}_j\rangle$ 
%% and $\langle \hat{F}_j \rangle_N$ 
always vanish as the system evolves.
We define the transverse magnetization as 
\begin{equation}
  \label{S_perp}
  S_\perp = \sqrt{\langle\hat{F}_x^2 +\hat{F}_y^2\rangle}.
\end{equation}
To examine whether the state converges to the fully magnetized steady state in Eq.~(\ref{rho_ss_general}) with Eqs.~(\ref{maxF_general}) and (\ref{rho_N_general}) 
(or with Eqs.~(\ref{states_F2N_31}) and (\ref{states_F2N_32}) for $q \neq 0$),
we introduce the normalized spin-squared operator as
\begin{equation}
  \label{F2norm}
\hat{\textbf{F}}^2_\mathrm{norm} = \frac{\hat{F}_x^2 + \hat{F}_y^2 + \hat{F}_z^2}{2\hat{N}(2\hat{N}+1)}.
\end{equation}
To avoid divergence arising from $N = 0$,
we define the density operator restricted to the $N > 0$ subspace as $\hat{\rho}_{N > 0}=\oplus_{N = 2}^{20} \hat{\rho}_N$.
The expectation value of Eq.~(\ref{F2norm}) is taken with respect to this restricted state as
\begin{equation}
F^2_\mathrm{norm} = \frac{\mathrm{Tr}[\hat{\textbf{F}}^2_\mathrm{norm} \hat{\rho}_{N > 0}]}{\mathrm{Tr}[\hat{\rho}_{N > 0}]}.
\end{equation}
This quantity becomes unity for the fully magnetized steady state in Eq.~(\ref{rho_ss_general}).
Let $P_N$ be the probability of finding the system in a state with $N$ particles, 
\begin{equation}
P_N = \frac{\mathrm{Tr}[\hat{\rho}_N]}{\mathrm{Tr}[\hat{\rho}]}.
\end{equation}
We define the probability of measuring a magnetization of $m$ within the $N$-particle subspace as 
\begin{equation}
Q_m = \frac{\mathrm{Tr}[\hat{\mathcal{P}}_m \hat{\rho}_N]}{\mathrm{Tr}[\hat{\rho}_N]}, 
\end{equation}
where $\hat{\mathcal{P}}_m$ is an operator describing the projection onto the eigenstates of $\hat{F}_z$ with an eigenvalue $m$.

%% Figure 1
\begin{figure}[!t]
\centering
\includegraphics{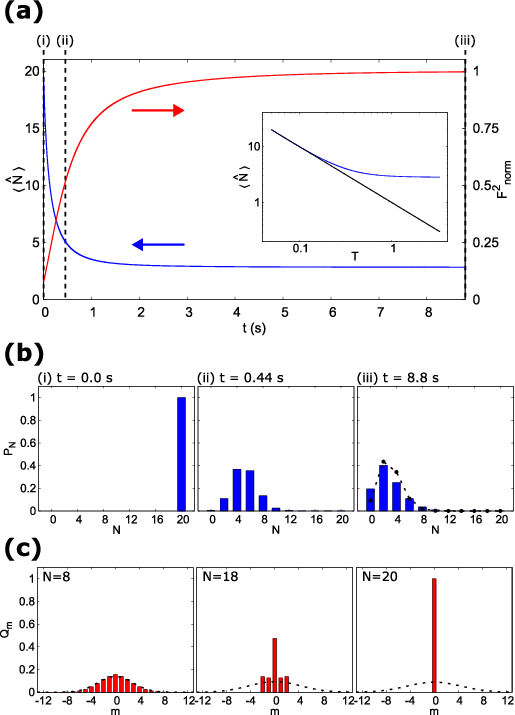}
\caption
{
Time evolution for $q = 0$ with the initial state in Eq.~(\ref{initial_state}).
(a) Average particle number $\langle \hat{N} \rangle$ (blue or dark gray line)
and normalized total magnetization $F^2_\mathrm{norm}$ (red or light gray line).
{\color{black}
Inset: Log-log plot of the average particle number $\langle \hat{N}
\rangle$ (blue or dark gray line) and
the expression in Eq.~(\ref{analytic_result}) (black line) as a function of the scaled time $T=bt/V^\mathrm{eff}+N_\mathrm{ini}^{-1}$ 
with $b = 0.0081 \times 10^{-11} \mathrm{cm}^3 / \mathrm{s}$.}
(b) Particle number distributions $P_N$ at (i) $t = 0$, (ii) $0.44~\mathrm{s}$, and (iii) $8.8~\mathrm{s}$. 
The dotted line in (iii) represents a Poisson distribution restricted to even integers in Eq.~(\ref{poisson_distribution}) with $\eta = 3.08$.
(c) Distributions $Q_m$ of the $z$-magnetization at $8.8~\mathrm{s}$. 
The dotted lines correspond to Eq.~(\ref{mean_field_coefficient_35.5}).
See Appendix for $Q_m$ in the subspaces with other $N$.
}\label{fig1}
\end{figure}

\subsection{Without quadratic Zeeman effect}
%% paragraph 1
First, we numerically solve the master equation for $q = 0$.
Figure~\ref{fig1}(a) shows the time evolution of the average particle number $\langle \hat{N} \rangle$ and the normalized total magnetization $F^2_\mathrm{norm}$.
The average particle number rapidly decreases in the early stage, 
since the initial density operator contains an abundance of states
with $F < 2N$, which are subject to inelastic decay.
After several hundred milliseconds, 
the states with $F < 2N$ are depleted and the decrease in $\langle \hat{N} \rangle$ becomes slow;
$\langle \hat{N} \rangle$ finally converges to $\simeq 3$ over long periods.
The normalized total magnetization $F^2_\mathrm{norm}$ increases monotonically, 
as shown in Fig.~\ref{fig1}(a), and finally converges to unity;
this behavior indicates that the final state is the fully magnetized steady state, given by Eq.~(\ref{rho_ss_general}).
Because of the rotational symmetry of the system,
the expectation values $\langle \hat{F}_x\rangle, \langle \hat{F}_y\rangle$, and $\langle \hat{F}_z\rangle$ of
the magnetization remain zero as the system evolves.

%% paragraph 2
{\color{black}
The decrease in $\langle \hat{N}\rangle$ in Fig.~\ref{fig1}(a) can roughly be estimated as follows.
Assuming that
$\mathrm{Tr} \left[\hat{A}_\mathcal{FM}^\dagger \hat{A}_\mathcal{FM}\hat{\rho} \right]$ 
in Eq.~(\ref{derivative_N}) is of the order of $\langle \hat{N}\rangle^2$, 
we have $d \langle \hat{N} \rangle / dt \sim -b \langle \hat{N} \rangle^2 / V^\mathrm{eff}$, where $b$ is a parameter comparable to the values of $b_0$ and $b_2$.
This differential equation is solved to give
\begin{equation}
\label{analytic_result}
  \langle \hat{N} \rangle = \frac{1}{\frac{b}{V^\mathrm{eff}} t + \langle \hat{N} \rangle_0^{-1}},
\end{equation}
where $\langle \hat{N} \rangle_0$ is the initial average particle number.
Fitting the numerical results for $\langle \hat{N} \rangle$ to Eq.~(\ref{analytic_result}) for $0 \leq t \leq 0.22~\mathrm{s}$ with $\langle \hat{N} \rangle_0 = N_\mathrm{ini}$, 
we find the value of $b$ to be $b \simeq 0.0081 \times 10^{-11} \mathrm{cm}^3 / \mathrm{s}$. 
The inset in Fig.~\ref{fig1}(a) compares Eq.~(\ref{analytic_result}) with the numerically obtained average particle number $\langle \hat{N} \rangle$ on a log-log scale, 
where $T = bt / V^\mathrm{eff} + N_\mathrm{ini}^{-1}$.
Equation~(\ref{analytic_result}) well describes the initial decrease in $\langle\hat{N} \rangle$.
The numerical result deviates from Eq.~(\ref{analytic_result}) for large $T$, 
since the fully magnetized atoms $\simeq 3$ remain in the steady state, which is not included in Eq.~(\ref{analytic_result}).
}

%% paragraph 3
{\color{black}
Figure~\ref{fig1}(b) displays the particle number distributions $P_N$ at different times.
As time progresses, we observe that the distribution becomes broad,
and that its peak moves toward smaller particle numbers because of inelastic loss.
For large $t$, the particle number distribution converges to a shape resembling a Poisson distribution restricted to even integers,
\begin{equation}
  \label{poisson_distribution}
P_N = \frac{\eta^N}{N!\cosh(\eta)},
\end{equation}
with a peak at $N = 2$.
Fitting the observed distribution to a Poisson distribution yields $\eta \simeq 3.08$, which is shown as the dotted line in Fig.~\ref{fig1}(b).
The mechanism that the observed particle number distribution resembles
a Poisson distribution merits further study.

%% paragraph 4
Since the initial state in Eq.~(\ref{initial_state}) is an eigenstate of $\hat{F}_z$ with an eigenvalue 0,
the initial magnetization distribution $Q_m$ in the subspace of $N = N_\mathrm{ini} = 20$ is localized at $m = 0$ (i.e., $Q_m = \delta_{m, 0}$).
At later times, $Q_m$ is broadened by inelastic loss of particles with $m_1 + m_2 \neq 0$, where $m_1$ and $m_2$ are the magnetic quantum numbers for the lost particles.
As a result, a statistical mixture of states with different $m$ is generated in the subspaces of $N < N_\mathrm{ini}$.
Figure~\ref{fig1}(c) shows $Q_m$ at $t = 8.8$ s, at which the distributions have almost converged to those of the final steady state.
The distribution $Q_m$ in the $N = N_\mathrm{ini} = 20$ subspace is still localized at $m = 0$, and is broader distributions in subspaces with smaller $N$.

%% paragraph 5
For comparison, we consider the $N$-particle fully polarized mean-field state,
\begin{equation}
\label{transverse_state}
|\Psi_\mathrm{MF}^{(N)}(\varphi)\rangle = \hat{R}(\varphi) |F = 2N, F_z = 2N\rangle,
\end{equation}
where $\hat{R}(\varphi) = e^{-i \varphi \hat{F}_x}e^{-i \pi / 2 \hat{F}_y} $ is the rotation operator.
This state is fully polarized in the $x$-$y$ direction with an
azimuthal angle $\varphi$, and all $N$ atoms occupy the same spin
state with a distribution
\begin{equation}
  \label{mean_field_coefficient_35.5}
  Q^\mathrm{MF}_m=\frac{1}{2^{4N}} \binom{4N}{2N+m}.
\end{equation}
The density operator after the time evolution is a statistical mixture
of different $m$ (not a superposition), since components with
different $m$ are generated by the jump term in the master equation,
not by the unitary part.
Using the mean-field states in Eq.~(\ref{transverse_state}) with different $\varphi$, we can construct such a statistical mixture as
\begin{equation}
\label{mean_field_like}  
\hat\rho_N \propto \int d\varphi |\Psi_\mathrm{MF}^{(N)}(\varphi)\rangle \langle \Psi_\mathrm{MF}^{(N)}(\varphi)|, 
\end{equation}
since the off-diagonal elements with different $m$ and $m'$ in the integrand are proportional to $e^{i(m'-m)\varphi}$ and vanish by integration.
The state in Eq.~(\ref{mean_field_like}) also has the same distribution as Eq.~(\ref{mean_field_coefficient_35.5}).

%% paragraph 6
The dotted lines in Fig.~\ref{fig1}(c) show the distribution in Eq.~(\ref{mean_field_coefficient_35.5}).
We find that $Q_m$ in the $N=8$ subspace (also in the subspaces with smaller $N$) agrees well with Eq.~(\ref{mean_field_coefficient_35.5}),
which indicates that the steady state reached after a long time is well approximated by a direct sum of Eq.~(\ref{mean_field_like}) with different $N$.
This result is consistent with the experimental and mean-field results in Ref.~\cite{Eto}, 
in which fully polarized states in the $x$-$y$ directions were obtained after time evolution.

%% paragraph 7
By contrast, as shown in Fig.~\ref{fig1}(c),
the distributions in the subspaces with larger particles ($N = 18$ and $20$) are narrower than Eq.~(\ref{mean_field_coefficient_35.5}).
In particular, for $N = N_\mathrm{ini} = 20$, the distribution $Q_m = \delta_{m, 0}$ remains unchanged from the initial state.
However, this state $\rho_{N=20}$ obtained as the steady state in the $N = 20$ subspace 
is strikingly different from the initial state in Eq.~(\ref{initial_state}).
This difference is easily understood from the simple two-particle example, 
in which the initial state in Eq.~(\ref{initial_state_expanded}) contains various states,
whereas the steady state in Eq.~(\ref{steady_state_2particles}) only contains the maximum spin state in the $N = N_\mathrm{ini}$ subspace.
The $N = N_\mathrm{ini} = 20$ subspace in the steady state should be
\begin{equation}
\label{rho_N20}
\rho_{N=20} \propto |F=2N, F_z=0\rangle \langle F=2N, F_z=0|, 
\end{equation}
according to the proof in Sec.~\ref{theory}B.
This state is highly nonclassical, since it can be written as
\begin{equation}
\label{dicke_pure}  
|F=2N, F_z = 0\rangle \propto \int d\varphi |\Psi_\mathrm{MF}^{(N)}(\varphi)\rangle, 
\end{equation}
where the terms with $F_z = m$ in the integrand are proportional to $e^{i m \varphi}$ and the $m \neq 0$ terms vanish by integration.
The state in Eq.~(\ref{dicke_pure}) is a superposition of the different mean-field states and can be regarded as a Schr\"odinger-cat-like state.
However, the probability of observing this state is very small: $P_{N=20} \sim 10^{-8}$ (see (iii) of Fig.~\ref{fig1}(b)).
}

%% Figure 3
\begin{figure}[!t]
\centering
\includegraphics{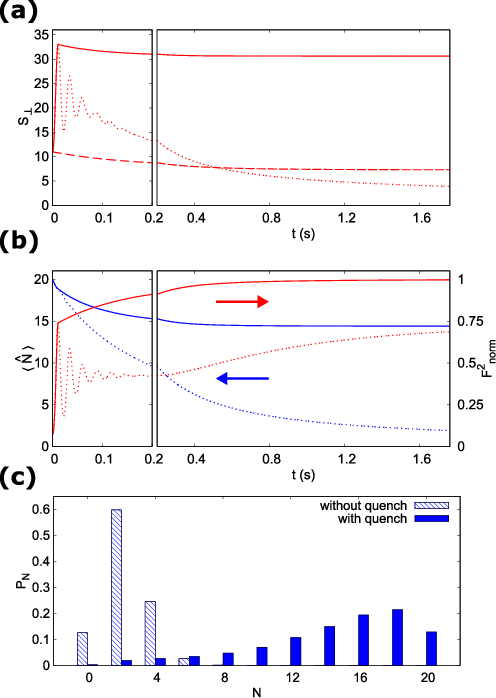}
\caption
{
Time evolution for $q / \hbar = 204~\mathrm{Hz}$ with and without
quenching of $q$.
In the case with quenching, $q$ is suddenly changed to 0 at $t = 11.0~\mathrm{ms}$.
(a) Transverse magnetization $S_\perp$ with quenching (solid line),
without quenching (dotted line), and without the quadratic Zeeman term
(dashed line).
(b) Average particle number $\langle \hat{N} \rangle$ (red or dark
gray lines) and normalized total magnetization $F^2_\mathrm{norm}$
(red or light gray lines)
with quenching (solid lines) and without quenching (dotted lines).
(c) Particle number distribution at $t = 1.76~\mathrm{s}$ with quenching (filled boxes) and without quenching (striped boxes).
}\label{fig3}
\end{figure}

\subsection{With quadratic Zeeman effect}
%% paragraph 1
{\color{black}
As shown in the previous subsection, the steady state in the $N=N_\mathrm{ini}=20$ subspace is the highly nonclassical state $|F=2N, F_z = 0\rangle$.
In general, deviations from the mean-field approximation become more significant in subspaces with larger $N$.
However, the survival probability $P_N$ for the steady state is extremely low
for large $N$, as shown in Fig.~\ref{fig1}(b), and experimental
observation is almost impossible.

%% paragraph 2
To enhance $P_N$ for the case with large $N$ in the steady state, we include the quadratic Zeeman effect.
As has been shown in Ref.~\cite{Eto}, the quadratic Zeeman effect changes the relative phases between different magnetic sublevels,
which can increase the transverse magnetization.
The increase in the transverse magnetization enhances $P_N$ for large $N$,
since the atomic spins tend to align in the same direction and atomic loss is suppressed.
}

%% paragraph 3
{\color{black}
We include the quadratic Zeeman term with $q / \hbar=204~\mathrm{Hz}$ in the Hamiltonian in Eq.~(\ref{HofSMA}), 
which corresponds to a magnetic field of $484~\mathrm{mG}$.
This value of $q$ is chosen so that the initial rise in the transverse magnetization is maximized.
}
The dotted line in Fig.~\ref{fig3}(a) shows the transverse magnetization $S_\perp$.
In the early stage of evolution, $S_\perp$ rapidly increases
because of the quadratic Zeeman effect, and then oscillates.
Such behavior has also been observed experimentally~\cite{Eto}.
Subsequently, $S_\perp$ decays to a small value because of the particle loss.
The dotted lines in Fig.~\ref{fig3}(b) show the time evolution of the average particle number $\langle \hat{N} \rangle$ and the normalized total magnetization $F^2_\mathrm{norm}$.
The value of $F^2_\mathrm{norm}$ first oscillates and then increases toward unity.
Although the final steady state should have $F^2_\mathrm{norm} = 1$ according to the proof in Sec.~\ref{theory}B, the convergence is slow.
The average particle number $\langle \hat{N} \rangle$ monotonically decreases with time
and converges to a small value, since the steady state consists of
only a limited set of states, as given by Eqs.~(\ref{states_F2N_31}) and (\ref{states_F2N_32}).
{\color{black}
Consequently, the probability $P_N$ is localized near $N = 2$, and $P_N$ for large $N$ is still very small in the steady state [Fig.~\ref{fig3}(c)].
In this scheme of a constant $q$, therefore, the subspaces with $N
\simeq N_\mathrm{ini}$ do not survive and a nonclassical state cannot be obtained.

%% paragraph 4
To obtain the nonclassical state, we quench the quadratic Zeeman coefficient to $q = 0$ at $t = 11.0~\mathrm{ms}$, at which $S_\perp$ reaches its first peak.
Quenching of $q$ is realized by switching off the external magnetic field.
The solid lines in Figs.~\ref{fig3}(a) and \ref{fig3}(b) show that
$\langle\hat{N}\rangle$ and $S_\perp$ remain large for a long time as
a result of quenching of $q$.
This is attributed to the survivable states being extended from Eqs.~(\ref{states_F2N_31}) and (\ref{states_F2N_32}) to Eq.~(\ref{maxF_general}).
Figure~\ref{fig3}(b) shows that $F^2_\mathrm{norm}$ converges to
unity, which indicates that the fully magnetized steady state is reached.
Figure~\ref{fig3}(c) shows that, in the steady state,
$P_N$ has a peak at $N = 18$ and the probability $P_{N_\mathrm{ini}} = P_{20}$ that no atoms are lost is $\simeq 0.13$,
which is much larger than that in the case without quenching of $q$.
Thus, the nonclassical state in Eq.~(\ref{rho_N20}) is obtained with a
probability of $13\%$ by this quench protocol.
As shown in Fig.~\ref{fig3}(c), the steady state consists not only of the $N = 20$ subspace but also of subspaces with other $N$.
If we can detect the lost atoms escaping from the trap,
the particle number $N$ remaining in the system can be determined and we obtain the pure nonclassical state in Eq.~(\ref{dicke_pure}).
}

\section{Conclusions}\label{conclusions}
We have investigated the dynamics of spin-2 bosons with atomic dissipation. 
In a previous related study~\cite{Eto},
it was shown experimentally and numerically that dissipation can
enhance the transverse magnetization in the time evolution, 
and the results of the mean-field calculations were found to agree
well with the experiments.
By contrast, the present work focused on the quantum many-body nature of a system with a small number of particles.
We proved that the steady states for this system are restricted to statistical mixtures of states with maximum total spins, 
given by Eq.~(\ref{rho_ss_general}) with Eqs.~(\ref{maxF_general}) and (\ref{rho_N_general}).
We performed numerical simulations using the master equation with and without the quadratic Zeeman term.
In the simulations without the quadratic Zeeman term, we confirmed that the system indeed
reached a steady state with the maximum total spins.
{\color{black}
However, the probability of obtaining the nonclassical state in Eq.~(\ref{rho_N20}) was extremely small.
In the simulations that included the quadratic Zeeman term, the transverse magnetization was enhanced, but the nonclassical state was still lost from the final steady state.}
When the quadratic Zeeman term was removed in the early stages of time
evolution, a relatively large number of particles was found to remain in the steady state.
{\color{black}
Therefore, the nonclassical state in Eq.~(\ref{rho_N20}) can be obtained with a high probability.
}
This finding suggests that particle dissipation not only destroys the quantum properties but also highlights nontrivial quantum many-body states.
Although we considered $^{87}\mathrm{Rb}$ atoms in the present study, other atomic species, such as $^{23}\mathrm{Na}$ atoms,
may be experimentally suitable, since their inelastic decay rates are much larger \cite{Gorlitz,Julienne,Burke} and steady states can be reached much faster.

\begin{acknowledgments}
This work was supported by JSPS KAKENHI Grant Number JP23K03276.
\end{acknowledgments}

\clearpage

\renewcommand{\thefigure}{\Alph{section}.\arabic{figure}}
\appendix*
  
{\color{black}
\section{Distribution $Q_m$ of $z$-magnetization in subspaces with different $N$}

\setcounter{figure}{0}
\begin{figure*}[!b]
\centering
\includegraphics{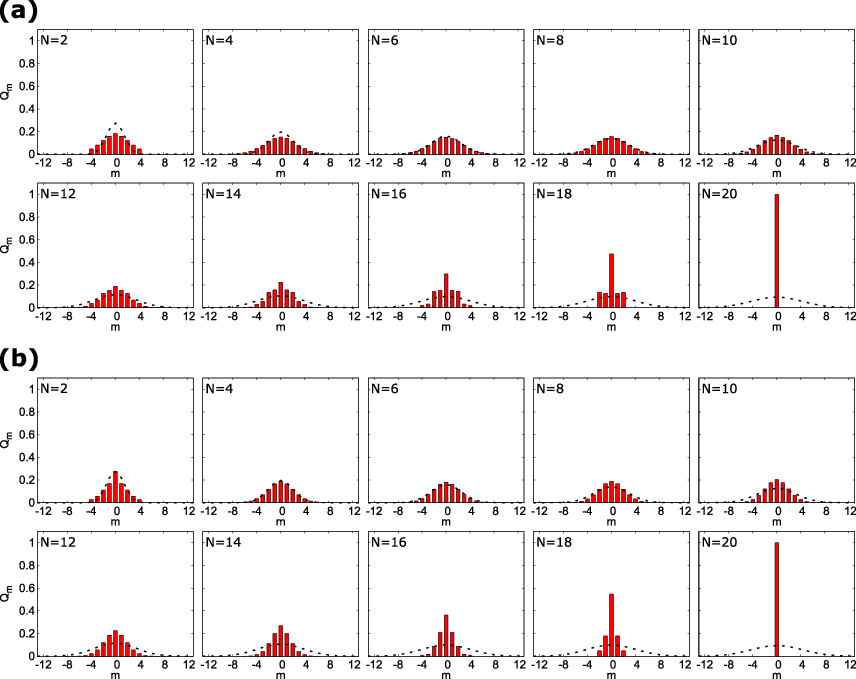}
\caption
{\color{black}
Distribution $Q_m$ of $z$-magnetization in subspaces with different particle numbers $N$ at final stage of evolution.
The dotted lines show Eq.~(\ref{mean_field_coefficient_35.5}).
(a) Distributions $Q_m$ for $q = 0$ at $t = 8.8~\mathrm{s}$, 
corresponding to Fig.~\ref{fig1} (panels for $N = 8$, 18, and 20 are the same as those in Fig.~\ref{fig1}(c)).
(b) Distributions $Q_m$ at $t = 1.76~\mathrm{s}$ with quenching from $q / \hbar = 204~\mathrm{Hz}$ to $0$ at $t = 11.0 \mathrm{ms}$, corresponding to Fig.~\ref{fig3}.
}\label{figA1}
\end{figure*}

In this Appendix, we present the $z$-magnetization distribution $Q_m$
in each subspace for different particle numbers $N$ at the final stage
of evolution.  
In Fig.~\ref{figA1}, the histograms represent the distributions $Q_m$ obtained from the numerical simulations, 
whereas the dotted lines plot Eq.~(\ref{mean_field_coefficient_35.5}).
Figure~\ref{figA1}(a) shows the results for $q = 0$ with the initial state in Eq.~(\ref{initial_state}).
The distributions agree well with Eq.~(\ref{mean_field_coefficient_35.5}) for small $N$ ($6 \leq N \leq 10$). 
By contrast, for large $N$, the distributions tend to localize around $m = 0$.
However, the probability of occupying these subspaces is very small, as discussed in the main text.
Figure~\ref{figA1}(b) shows the distributions after quenching from $q / \hbar = 204~\mathrm{Hz}$ to $q = 0$. 
The distributions for small $N$ ($4 \leq N \leq 8$) are well described by Eq.~(\ref{mean_field_coefficient_35.5}).
As in the $q = 0$ case, the distributions for large $N$ are localized around $m = 0$;
however, they appear with substantially higher probabilities than those for $q = 0$ (see Figs.~\ref{fig1}(b) and \ref{fig3}(b)).
In the $N = N_{\rm ini} = 20$ subspace, the states are strictly restricted to the $F_z = 0$ sector.
These states are described by Eqs.~(\ref{rho_N20}) and (\ref{dicke_pure}).
}

\clearpage

\begin{thebibliography}{99}
  \bibitem{Kawaguchi} Y. Kawaguchi and M. Ueda, Spinor Bose-Einstein condensates, Phys. Rep. \textbf{520}, 253 (2012).
  \bibitem{Stamper-Kurn} D.M. Stamper-Kurn, M.R. Andrews, A.P. Chikkatur, S. Inouye, H.J. Miesner, J. Stenger, and W. Ketterle, Optical Confinement of a Bose-Einstein Condensate, Phys. Rev. Lett. \textbf{80}, 2027 (1998).
  \bibitem{Stenger} J. Stenger, S. Inouye, D.M. Stamper-Kurn, H. J. Miesner, A.P. Chikkatur, and W. Ketterle, Spin domains in ground-state Bose-Einstein condensates, Nature (London) \textbf{396}, 345 (1998).
  \bibitem{Zhang} W. Zhang, S. Yi, and L. You, Mean field ground state of a spin-1 condensate in a magnetic field, New J. Phys. \textbf{5}, 77 (2003).
  \bibitem{Murata} K. Murata, H. Saito, and  M. Ueda, Broken-axisymmetry phase of a spin-1 ferromagnetic Bose-Einstein condensate, Phys. Rev. A \textbf{75}, 013607 (2007).
  \bibitem{Leanhardt} A. E. Leanhardt, A. Görlitz, A. P. Chikkatur, D. Kielpinski, Y. I. Shin, D. E. Pritchard, and W. Ketterle, Imprinting vortices in a Bose-Einstein Condensate using Topological phases, Phys. Rev. Lett. \textbf{89}, 190403 (2002).
  \bibitem{Kumakura} M. Kumakura, T. Hirotani,  M. Okano, Y. Takahashi, and T. Yabuzaki, Topological formation of a multiply charged vortex in the Rb Bose-Einstein condensate: Effectiveness of the gravity compensation, Phys. Rev. A \textbf{73}, 063605 (2006). 
  \bibitem{Leslie} L. S. Leslie, A. Hansen, K. C. Wright, B. M. Deutsch, and N. P. Bigelow, Creation and Detection of Skyrmions in a Bose-Einstein Condensate, Phys. Rev. Lett. \textbf{103}, 250401 (2009).
  \bibitem{Ray1} M. W. Ray, E. Ruokokoski, S. Kandel, M. M\"{o}tt\"{o}nen, and D. S. Hall, Observation of Dirac monopoles in a synthetic magnetic field, Nature (London) \textbf{505}, 657 (2014). 
  \bibitem{Ray2} M. W. Ray, E. Ruokokoski, K. Tiurev, M. M\"{o}tt\"{o}nen, and D. S. Hall, Observation of isolated monopoles in a quantum field, Science \textbf{348}, 6234 (2015). 
  \bibitem{Hall} D. S. Hall, M. W. Ray, K. Tiurev, E. Ruokokoski, A. H. Gheorghe, and M. M\"{o}tt\"{o}nen, Tying quantum knots, Nat. Phys. \textbf{12}, 478 (2016).
  \bibitem{Lee} W. Lee, A. H. Gheorghe, K. Tiurev, T. Ollikainen, M. M\"{o}tt\"{o}nen, and D. S. Hall, Synthetic electromagnetic knot in a three-dimensional skyrmion, Sci. Adv. \textbf{4}, eaao3820 (2018).
  \bibitem{Ciobanu} C. V. Ciobanu, S.K. Yip, and T.-L. Ho, Phase diagrams of $F = 2$ spinor Bose-Einstein condensates, Phys. Rev. A \textbf{61}, 033607 (2000). 
  \bibitem{Klausen} N. N. Klausen, J. L. Bohn, and C. H. Greene, Nature of spinor Bose-Einstein condensates in rubidium, Phys. Rev. A \textbf{64}, 053602 (2001).
  \bibitem{Widera} A. Widera, F. Gerbier, S. Fölling, T. Gericke, O. Mandel, and I. Bloch, Precision measurement of spin-dependent interaction strengths for spin-1 and spin-2 $^{87} \mathrm{Rb}$ atoms, New J. Phys. \textbf{8}, 152 (2006).
  \bibitem{Kuwamoto} T. Kuwamoto, K. Araki, T. Eno, and T. Hirano, Magnetic field dependence of the dynamics of $^{87} \mathrm{Rb}$ spin-2 Bose-Einstein condensates, Phys. Rev. A \textbf{69}, 063604 (2004).
  \bibitem{Schmaljohann} H. Schmaljohann, M. Erhard, J. Kronjäger, M. Kottke, S. van Staa, L. Cacciapuoti, J. J. Arlt, K. Bongs, and K. Sengstock, Dynamics of $F = 2$ Spinor Bose-Einstein Condensates, Phys. Rev. Lett. \textbf{92}, 040402 (2004).
  \bibitem{Saito} H. Saito and M. Ueda, Diagnostics for the ground-state phase of a spin-2 Bose-Einstein condensate, Phys. Rev. A \textbf{72}, 053628 (2005).
  \bibitem{Tojo1} S. Tojo, A. Tomiyama, M. Iwata, T. Kuwamoto, and T. Hirano, Collision dynamics between stretched states of spin-2 $^{87} \mathrm{Rb}$ Bose-Einstein condensates, Appl. Phys. B: Lasers Opt. \textbf{93}, 403 (2008).
  \bibitem{Tojo2} S. Tojo, T. Hayashi, T. Tanabe, T. Hirano, Y. Kawaguchi, H. Saito, and M. Ueda, Spin-dependent inelastic collisions in spin-2 Bose-Einstein condensates, Phys. Rev. A \textbf{80}, 042704 (2009).
  \bibitem{Eto} Y. Eto,  H. Shibayama, K. Shibata, A. Torii, K. Nabeta, H. Saito, and T. Hirano, Dissipation-Assisted Coherence Formation in a Spinor Quantum Gas, Phys. Rev. Lett. \textbf{122}, 245301 (2019).
  \bibitem{Koashi} M. Koashi and M. Ueda, Exact Eigenstates and Magnetic Response of Spin-1 and Spin-2 Bose-Einstein Condensates, Phys. Rev. Lett. \textbf{84}, 1066 (2000).
  \bibitem{Ueda} M. Ueda and M. Koashi, Theory of spin-2 Bose-Einstein condensates: Spin correlations, magnetic response, and excitation spectra, Phys. Rev. A \textbf{65}, 063602 (2002).
  \bibitem{Mueller} E. J. Mueller, T.-L. Ho, M. Ueda, and G. Baym, Fragmentation of Bose-Einstein condensates, Phys. Rev. A \textbf{74}, 033612 (2006).
  \bibitem{Evrard1} B. Evrard, A. Qu, J. Dalibard, and F. Gerbier, Observation of fragmentation of a spinor Bose-Einstein condensate, Science \textbf{373}, 1340 (2021).
  \bibitem{Evrard2} B. Evrard, A. Qu, J. Dalibard, and F. Gerbier, From Many-Body Oscillations to Thermalization in an Isolated Spinor Gas, Phys. Rev. Lett. \textbf{126}, 063401 (2021).
  \bibitem{Gardiner} C. W. Gardiner and P. Zoller, \textit{Quantum Noise} (Springer, Berlin, 2000).
  \bibitem{Gorlitz} A. G\"{o}rlitz, T. L. Gustavson, A. E. Leanhardt, R. L\"{o}w, A. P. Chikkatur, S. Gupta, S. Inouye, D. E. Pritchard, and W. Ketterle, Sodium Bose-Einstein Condensates in the $F = 2$ State in a Large-Volume Optical Trap, Phys. Rev. Lett. \textbf{90}, 090401 (2003).
  \bibitem{Julienne} P. S. Julienne, F. H. Mies, E. Tiesinga, and C. J. Williams, Collisional Stability of Double Bose Condensates, Phys. Rev. Lett. \textbf{78}, 1880 (1997).
  \bibitem{Burke} J. P. Burke, J. L. Bohn, B. D. Esry, and C. H. Greene, Impact of the $^{87}\mathrm{Rb}$ singlet scattering length on suppressing inelastic collisions, Phys. Rev. A \textbf{55}, R2511 (1997).
\end{thebibliography}
\end{document}